\documentclass[preprintnumbers,floatfix,nofootinbib]{revtex4}

\pdfoutput=1 % if your are submitting a pdflatex to arXIv 
\usepackage[tbtags]{amsmath}  % AMS math 
\usepackage{amssymb}          % AMS symbols
\usepackage{bm}               % bold math
\usepackage{graphicx}         % PostScript figures
\usepackage[export]{adjustbox}% to align images
\usepackage{hhline,multirow}  % for nicer tables
\usepackage{dcolumn}          % Align table columns on decimal point
\usepackage{slashed}
\usepackage{datetime}
\usepackage{color}
\usepackage[dvipsnames]{xcolor}

\newcommand{\lslash}{l \hspace{-0.19cm} / \,}
\newcommand{\nslash}{n \hspace{-0.24cm} / \,}

\newcommand{\xbj}{{x_B}}                   % Bjorken variable
\newcommand{\de}{d}                    % differential
\newcommand{\ii}{i}                    % imaginary unit

\newcommand{\eg}{{\em e.g.}}
\newcommand{\mj}{M_q}
\newcommand{\mq}{m_q}

\newcommand{\id}{{\bm 1}
}

\allowdisplaybreaks[2]

\begin{document}

%%%%%%%%%%%%%%%%%%%%%%%%%%%%%%%%%%%%%%%%%%%%%%%%%%%%%%%%%%%%%%%%%%%%%%
%%%%%%%%%%%%%%%%%%%%%%%   TITLE    %%%%%%%%%%%%%%%%%%%%%%%%%%%%%%%%%%%
%%%%%%%%%%%%%%%%%%%%%%%%%%%%%%%%%%%%%%%%%%%%%%%%%%%%%%%%%%%%%%%%%%%%%%

\preprint{JLAB-THY-17-2483}

\title{Accessing the nucleon transverse structure in inclusive deep inelastic scattering} 

\author{Alberto~Accardi$^{a}$, Alessandro~Bacchetta$^{b}$} 
\affiliation{
$^a$Hampton University, Hampton, VA 23668, USA, and Jefferson Lab, Newport News, VA 23606, USA \\
$^b$Dipartimento di Fisica, Universit\`a degli Studi di Pavia, and INFN,
Sez. di Pavia, 27100 Pavia, Italy
}

%\date{\bf Thursday, June 6th, 2017}
%\date{\bf \today, \currenttime}

\begin{abstract}
We revisit the standard analysis of inclusive Deep Inelastic Scattering
off nucleons taking into  account the fact that on-shell quarks cannot be present in the final state, but they rather decay into hadrons --  a process that can be described in terms of suitable ``jet'' correlators. As a consequence, a spin-flip term
associated with the invariant mass of the produced hadrons is generated
nonperturbatively and couples to the target's transversity distribution
function. In inclusive cross sections, this provides an hitherto neglected and
large contribution to the twist-3 part of the $g_2$ structure function, that
can explain the discrepancy between recent calculations and fits of this
quantity.  It also provides an extension of the Burkhardt--Cottingham sum rule, 
providing new information on the transversity function, as well as an extension of the Efremov--Teryaev--Leader sum rule, suggesting a novel way to measure the tensor charge of the proton.
\end{abstract}

%\pacs{ }

%\keywords{ }

\maketitle

%%%%%%%%%%%%%%%%%%%%%%%%%%%%%%%%%%%%%%%%%%%%%%%%%%%%%%%%%%%%%%%%%%%%%%%%%
\section{Introduction}

The tensor charge is a fundamental property of the nucleon that is at
present poorly constrained but of fundamental importance, not the least because its knowledge can also be used to put constraints on searches for physics beyond the Standard Model~\cite{Cirigliano:2013xha,Bhattacharya:2015esa,Courtoy:2015haa,Tomalak:2017owk,Yamanaka:2017mef}.
The tensor charge has been estimated in lattice QCD (see, \eg,
\cite{Green:2012ej,Bali:2014nma,Bhattacharya:2015wna,Abdel-Rehim:2015owa,Bhattacharya:2016zcn}), but
only limited information is available from direct measurements. Its experimental extraction requires first of all flavor-separated measurements of the so-called transversity parton distribution function,
denoted by $h_1^q(x)$ (see Ref.~\cite{Barone:2001sp} for a review and
Refs.~\cite{Radici:2015mwa,Anselmino:2015sxa,Kang:2015msa} for the most recent
extractions). Secondly, one needs to perform flavor-by-flavor integrals of these, that correspond to the contribution of a parton flavor $q$ to the tensor charge. 

The transversity distribution is notoriously difficult to access because it is
a chiral-odd function and needs to be combined with a spin-flip mechanism to
appear in a scattering process~\cite{Jaffe:1996zw}. Usually, this spin flip is provided by another
nonperturbative distribution or fragmentation function, accessible in Drell-Yan or semi-inclusive Deep Inelastic Scattering (DIS)~\cite{Ralston:1979ys,Jaffe:1991kp,Jaffe:1993xb,Collins:1992kk}.
The only other known way to attain spin-flip terms in Quantum Electro-Dynamics and QCD is taking into account mass corrections. In fact, it is well known that the transversity distribution gives a contribution to the structure function $g_2$ in inclusive DIS (see, \eg, \cite{Accardi:2009au} and references therein), and in particular to the violation of the so-called Wandzura--Wilczek relation for $g_2$~\cite{Wandzura:1977qf}. However, this contribution is proportional to the current quark mass and can be expected to be negligibly small.

In this paper, we discuss a novel way of accessing the transversity parton distribution function (PDF) and measuring the proton's tensor charge in totally inclusive Deep Inelastic Scattering. 
We revisit the standard analysis of the DIS handbag diagram, taking into account the fact that on-shell quarks cannot, in fact, be present in the final state, but they rather decay and form (mini)jets of hadrons. This is sufficient to modify the structure of the DIS cut diagram, even if none of those hadrons is detected in the final state. For a proper description of this effect, we include ``jet correlators'' into the analysis, and pay particular attention to ensuring that our results are gauge invariant.

The jet correlators describe interactions of a perturbative quark with vacuum fields, that break chiral symmetry and generate a nonperturbative mass
estimated in the 10-100 MeV range, potentially
much larger than the current quark mass for light flavors, as also heuristically advocated in Ref.~\cite{Afanasev:2007ii} for a study of transverse target single-spin asymmetries in two-photon exchange processes. Here, we formalize this idea in the context of collinear factorization, and observe that jet correlators introduce a new
contribution already in one-photon exchange processes, and more precisely to the inclusive $g_2$ structure function. The new term is proportional to the transversity distribution function multiplied by a new nonperturbative ``jet mass'', which will be precisely defined below, and has the interesting features that:
(a) it violates the Wandzura--Wilczek relation;
(b) it extends the Burkhardt--Cottingham sum rule, providing new useful information on behavior of the transversity distribution;
(c) it also extends the Efremov--Teryaev--Leader sum rule, providing a novel way to measure the proton's tensor charge.
We estimate this new jet-mass-induced contribution based on a recent extraction of the transversity distribution, and show it can indeed be very large.

%%%%%%%%%%%%%%%%%%%%%%%%%%%%%%%%%%%%%%%%%%%%%%%%%%%%%%%%%%%%%%%%%%%%%%%%%
%\section{Jet correlator and twist-2 structure functions}
\section{The quark-quark jet correlator}

\begin{figure*}[bt]
  \centering
  (a)\includegraphics[width=0.3\linewidth,valign=t]{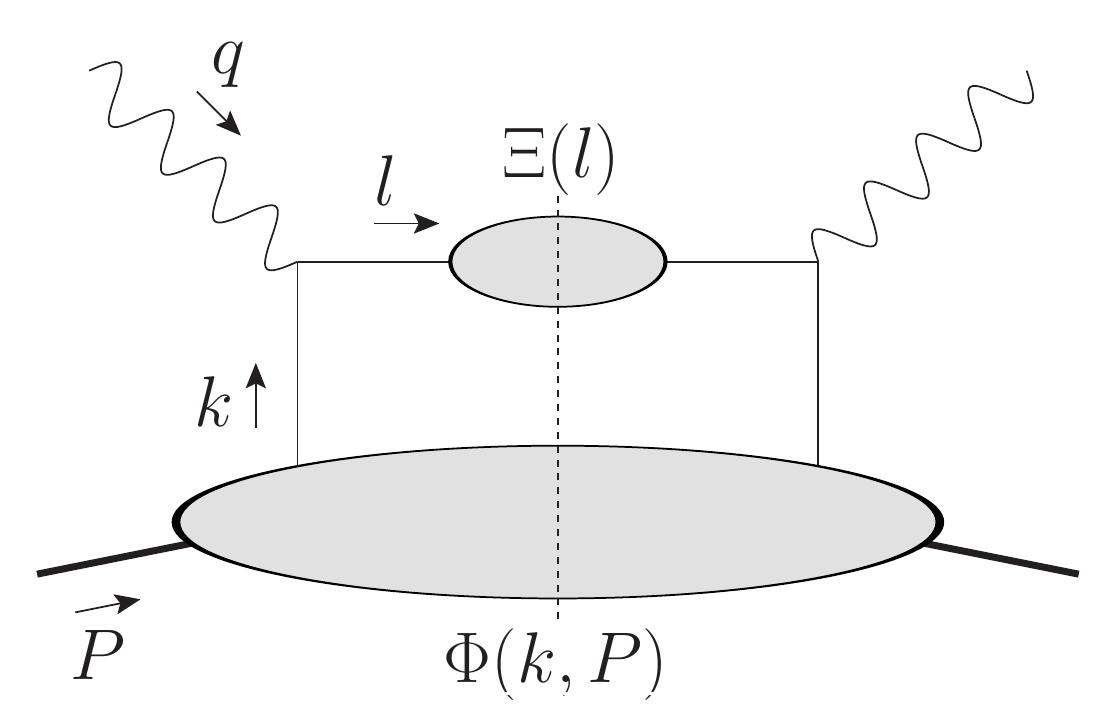}
  \hfill
  (b)\includegraphics[width=0.3\linewidth,valign=t]{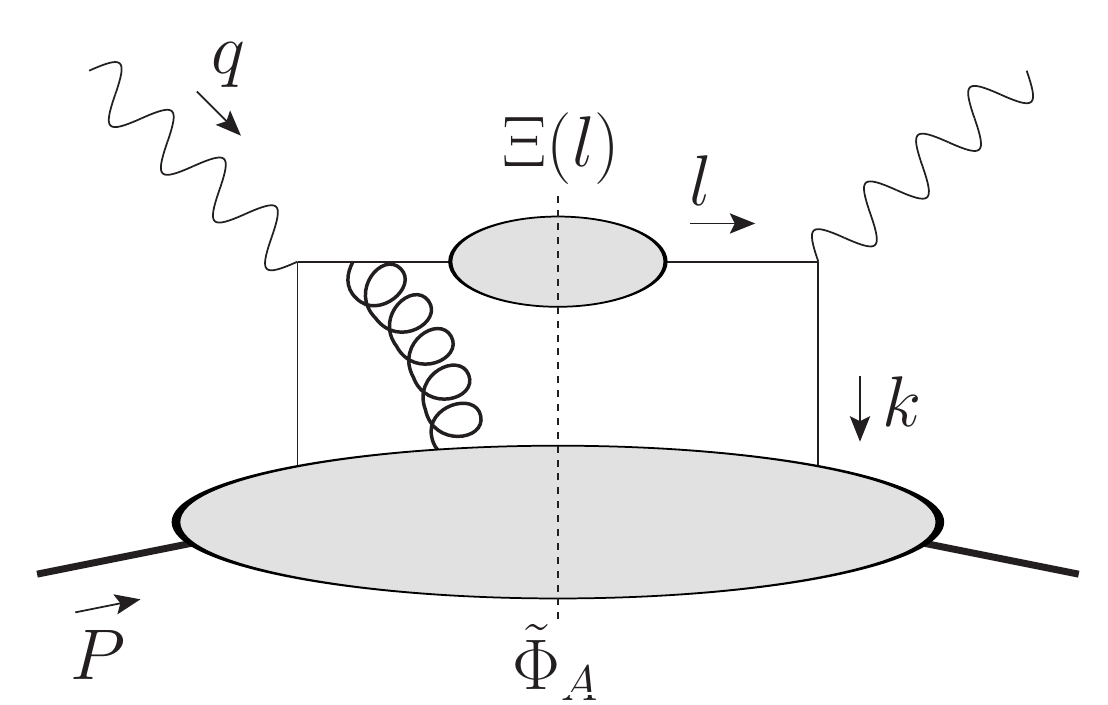}
  \hfill
  (c)\includegraphics[width=0.3\linewidth,valign=t]{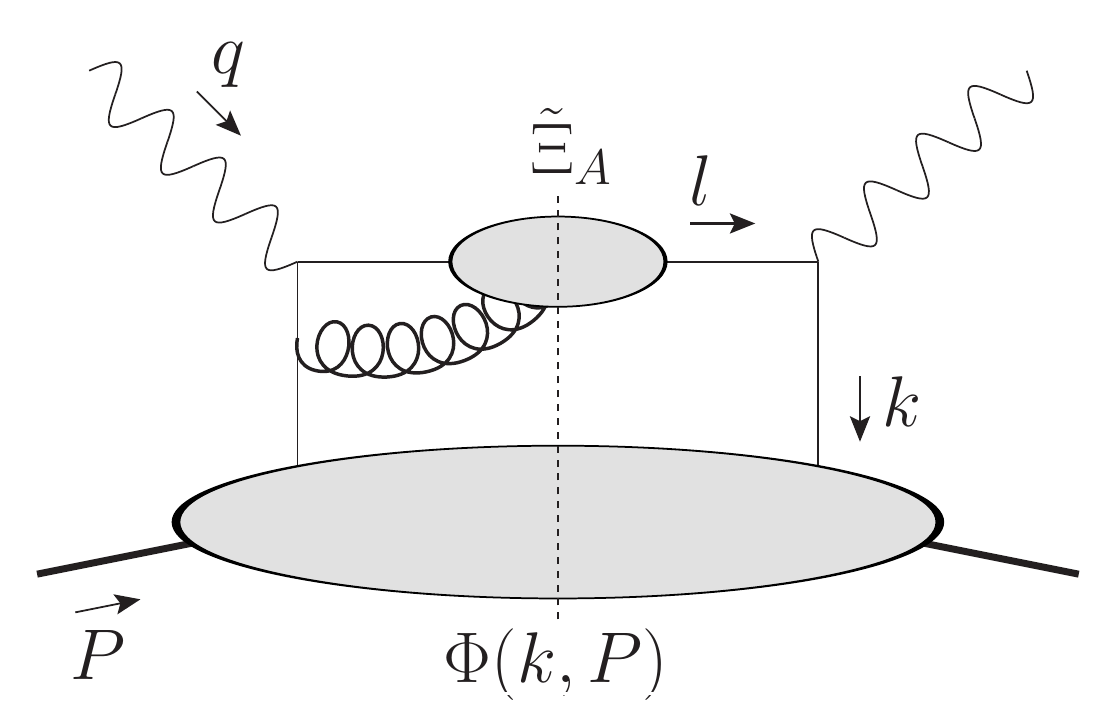}
  \caption{Diagrams contributing to inclusive DIS scattering up to twist-3, including a jet correlator in the top part. The proton is moving dominantly in the light-cone plus direction, and the jet in the minus direction. In diagrams (b) and (c), the gluon attaches to both the nucleon and jet correlators. The Hermitian conjugates of these two diagrams, i.e., with gluons attaching to the right of the cut, are not shown.
  }
  \label{fig:handbags}
\end{figure*}

Motivated by mass corrections to inclusive DIS structure functions at large values of the Bjorken invariant $x_B$,
Accardi and Qiu \cite{Accardi:2008ne} have introduced in the LO handbag diagram 
a ``jet correlator'', also called ``jet factor'' by Collins, Rogers, and Stasto in Ref.~\cite{Collins:2007ph}, that accounts for invariant mass production in the current region and ensures that leading twist calculations in collinear factorization are consistent with the $x_B<1$ requirement imposed by baryon number conservation. \cite{Accardi:2008ne}. The jet correlator is depicted in
Figure~\ref{fig:handbags}(a) and is defined as 
\begin{align}
\Xi_{ij}(l,n_+) =\int
  \frac{\de^4\eta}{(2\pi)^4}\; e^{i l \cdot \eta}\,
    \langle 0|\, {\cal U}^{n_+}_{(+\infty,\eta)}
\,\psi_i(\eta)
             \bar{\psi}_j(0)\,
{\cal U}^{n_+}_{(0,+\infty)}\,   |0 \rangle \ ,
\label{e:xifull}
\end{align} 
In this definition, $l$ is the quark's four-momentum, $\Psi$ the quark field
operator (with quark flavor index omitted for simplicity), and $|0\rangle$ is
the nonperturbative vacuum state. Furthermore, the correlator's gauge invariance is explicitly guaranteed the two Wilson line operators ${\cal
  U}^{n_+}$, that run to infinity first along along a light-cone plus direction determined by the vector $n_+$, then along the direction transverse to that vector, see \cite{Bacchetta:2006tn} for details. This path choice for the Wilson line is required by QCD factorization theorems, and the vector $n_+$ is determined by the particular hard process to which the jet correlator contributes. For example, in the case of inclusive DIS
discussed in this paper, this is determined by the four momentum transfer $q$
and the proton's momentum $p$. 

The correlator $\Xi$ can be parametrized in terms of jet parton correlation
functions $A_i$ and $B_i$ through a Lorentz covariant Dirac decomposition that utilizes the vectors $l$ and $n_+$, 
\begin{equation}
\Xi(l,n_+) = \Lambda A_1(l^2)\,\id + A_2(l^2)\,\lslash 
+ \frac{\Lambda^2}{l \cdot n_{+}} \nslash_+ \, B_{1}(l^2)
+ \frac{i \Lambda}{2 l \cdot n_{+}} [\,\lslash,\nslash_+ ] \, B_{2}(l^2) \ ,
\label{e:jetexpansion}
\end{equation} 
where $\Lambda$ is an arbitrary scale, introduced for power counting purposes.
In this parametrization, no terms proportional to $\gamma_5$ enter because of parity invariance. Time reversal invariance in QCD requires $B_{2}=0$, while $B_{1}$ contributes only at twist-4 order and will not be considered further in this paper. We focus, instead, on the role of chiral odd terms in the $g_2$ structure function up to twist 3. At this order, 
\begin{equation}
  \Xi(l,n_+) = \Lambda A_1(l^2)\,\id + A_2(l^2)\,\lslash 
    + {\cal O}(\Lambda^2/Q^2)
\label{e:jetexpansion-tw3}
\end{equation} 
is nothing else than the cut quark propagator; note however, that we consider
here the full QCD vacuum rather than the perturbative one (or, in other words, the interacting rather than the free quark fields).
The $A_1$ and $A_2$ terms can be interpreted in terms of the spectral
representation of the cut quark propagator (see, e.g., Sec.~6.3 of
\cite{D'Hoker:2004aa} and Sec.~2.7.2 of \cite{Romao:2013aa}),
\begin{align} 
  \Xi(l) =  
  \int d \sigma^2 \big[ J_1 (\sigma^2)\,\sigma\,\id + J_2 (\sigma^2)\,\lslash \big] \,
  \delta(l^2 -\sigma^2) \ ,
\label{e:jetspectral}
\end{align}
where $\sigma^2$ can be interpreted as the invariant mass of the current jet, {\it
  i.e.}, of the particles going through the cut in the top blob of
Fig.\ref{fig:handbags}(a). The $J_i$ are the spectral functions of the
quark propagator, also called ``jet functions'' in
\cite{Accardi:2008ne}, and can be interpreted as current-jet mass distributions. As a consequence of positivity constraints and CPT invariance, these satisfy \cite{D'Hoker:2004aa,Romao:2013aa,Weinberg:1995mt}
\begin{align}
  J_2(\sigma^2) \geq J_1(\sigma^2) \geq 0
  \hspace*{0.5cm} \text{and} \hspace*{0.5cm}
  \int d\sigma^2 J_2(\sigma^2) = 1 \ .
\label{eq:jetfnsprops}
\end{align}
From a comparison of Eqs.\eqref{e:jetexpansion} and \eqref{e:jetspectral}, one can see that 
\begin{align}
  A_1(l^2)&=\frac{\sqrt{l^2}}{\Lambda}J_1(l^2) & A_2(l^2)&=J_2(l^2) \ .
  \label{eq:jet_vs_spectral}
\end{align}

When inserting the jet correlator in the handbag diagram for inclusive DIS, the integration over $dl^+$, or equivalently $dl^2/(2l^-)$, is kinematically coupled to the other integrations, and induces corrections of order ${\cal O}(1/Q^2)$ whose effect on the $F_2$ structure function has been studied in Ref.~\cite{Accardi:2008ne}. In this paper, where we limit our
attention to effects of order ${\cal O}(1/Q)$, we can neglect $k^-$ compared to $q^-$. As a consequence, we can extend the integration over $dl^2$ to infinity, with the consequence that the jet correlator decouples from the parton correlator $\Phi$, and the inclusive structure functions only depend on the integrated jet correlator 
\begin{equation} 
  \Xi(l^-,{\bm l_T}) \equiv \int \frac{dl^2}{2l^-} \, \Xi(l) 
    =  \frac{\Lambda}{2 l^-}\,\xi_1 \id
    +  \xi_2 \frac{\nslash_-}{2} 
    + {\cal O}(l_T/l^-) + \text{higher\ twists} \ .
\end{equation} 
The neglected ${\bm l_T}$-dependent and higher twist terms only contribute to ${\cal O}(1/Q^2)$ to the inclusive cross section. Note that thanks to Eq.~\eqref{eq:jetfnsprops} we obtain
\begin{align}
\xi_1 &= \int d\sigma^2 \frac{\sigma}{\Lambda} J_1(\sigma^2) 
       \equiv \frac{\mj}{\Lambda},
&
\xi_2 &= \int d\sigma^2 J_2(\sigma^2) = 1 \ ,
\end{align} 
where $\mj$ can be interpreted as the average invariant mass produced in the spin-flip fragmentation processes of a quark of flavor $q$.

It is important to notice that, while $\xi_2=1$ exactly due to CPT invariance
(see Sec.~10.7 of Ref.~\cite{Weinberg:1995mt}), the jet mass
$\mj < \int d\sigma^2 \sigma \, J_2(\sigma^2)$ is dynamically determined. From the analytic properties of the spectral functions we expect that $J_2(\sigma^2) = Z \delta(\sigma^2-m_q^2) + \bar J_2(\sigma^2) \theta (\sigma^2-m_\pi^2)$, with $Z<1$ and the continuum starting at $m_\pi$ (the mass of the pion) due to color confinement effects, indicating $M_q=O(\Lambda_{QCD})$. However, in a dynamical confinement scenario, the spectral function $J_2$ needs not be positive definite \cite{Roberts:2016vyn} and we therefore estimate
$
  M_q\sim 10-100 \text{\ MeV}.
$
An experimental measurement of $M_q$ is anyway possible, as we discuss in Section~\ref{sec:moments}, and could shed some light on the confinement mechanism. We have also explicitly verified that $M_q>m_q$ in a model where quark fragmentation is simulated by a Yukawa pseudoscalar quark-meson interaction, already utilized, {\it e.g.}, in Ref.~\cite{Bacchetta:2001di}.

Although $\mj$ is in general a nonperturbative quantity, it is interesting to
notice that on the perturbative vacuum
\begin{align}
  \Xi^{\text{pert}}(l) = (\lslash + \mq \id) \, \delta(l^2-m_q^2) + {\cal O}(\alpha_s) \ ,
\end{align}
where $\mq$ is the current quark mass; therefore $\mj^\text{pert}=m_q$, and one recovers the result of the calculation with the conventional handbag diagram.
However, we are here considering nonperturbative effects
in the quark propagation, and $\mj \gg \mq$. 
Therefore, differently from $J_2$, the $J_1$ function leaves an imprint on the inclusive DIS cross section even in the asymptotic $Q^2 \to \infty$ regime.

\section{Twist-3 analysis}

Extending this analysis to the calculation of twist-3
structure functions requires not only to consider the $\xi_1$ term in the jet
correlator, but also quark-gluon-quark correlators in both the proton and the
vacuum as depicted in Figs.\ref{fig:handbags}(b) and (c), respectively. 

In diagram (b), the $\xi_1$ term contributes to ${\cal O}(1/Q^2)$, so that up to ${\cal O}(1/Q)$ considered in this paper this give the same contribution as in the conventional handbag calculation. The novel element in our analysis, instead, is the jet's quark-gluon-quark correlator $\Xi_A^{\mu}(l,k)$ in diagram (c), defined as 
\begin{equation} 
\begin{split} 
  \left(\tilde \Xi_A^{\mu} \right)_{ij} &=
   \int \frac{\de^4 \eta}{(2\pi)^{4}}\;
   e^{\ii k \cdot \eta}\,
   \langle 0|\,
   {\cal U}^{n_+}_{(+\infty,\eta)}\,
   g A^{\mu}(\eta)\,
   \,\psi_i(\eta)
   \,\bar{\psi}_j(0)\,
   {\cal U}^{n_+}_{(0,+\infty)}
   |0\rangle \ .
\label{e:xi_A}
\end{split} 
\end{equation}  
This diagram and its Hermitian conjugate are not only important to account for
all contribution of order ${\cal O}(1/Q)$, but also to restore gauge invariance, which is broken in diagram \ref{fig:handbags}(a) due to the different mass
of the incoming and outgoing quark lines, namely, $\mq \neq \mj$. 

Rather than directly using the definition \eqref{e:xi_A}, it is convenient and instructive to calculate the inclusive cross section as an integral of the semi-inclusive one summed over all produced hadron flavors, then utilize the QCD equations of motion, sum over all hadron flavors, and take advantage of  
\begin{align}
  \label{eq:SIDIS_to_DIS}
  \sum_h \int d^2p_{hT}\frac{dp_h^-}{2p_h^-} \,p_h^- \, \Delta^h(l,p_h) = l^-\,\Xi(l) \ , 
\end{align}
where $\Delta^h$ is the quark fragmentation correlator for production of a
hadron of flavor $h$ and momentum $p_h$, discussed in detail in Ref.~\cite{Bacchetta:2006tn}. In terms of
the TMD fragmentation functions we are interested in, the sum rule \eqref{eq:SIDIS_to_DIS} reads 
\begin{align}
  \label{eq:SIDIS_to_DIS_TMDlevel1}
  \sum_h \int dz d^2p_{hT} z D_1^h(z,p_{hT}) & = \xi_2 = 1   \\
  \label{eq:SIDIS_to_DIS_TMDlevel2}
  \sum_h \int dz d^2p_{hT} E^h(z,p_{hT}) & = \xi_1 = \mj / \Lambda\ ,
\end{align}
where $D_1^h(z,p_{hT})$ is the twist-2 quark fragmentation function, that depends on the hadron's collinear momentum fraction $z$ and transverse
momentum $p_{hT}$, and $E^h(z,p_{hT})$ is a chiral-odd twist-3 function
defined in \cite{Bacchetta:2006tn}. 

The role of the $\xi_1=M_q/\Lambda$ term in inclusive DIS can be discussed by analyzing the following terms of the semi-inclusive hadronic tensor \cite{Mulders:1995dh}:
\begin{align}
  \label{eq:Wsidis_ini}
  2 M  W^{\mu\nu}
    & = i \frac{2M}{Q} \hat t^{[\mu}_{\phantom \perp} 
    \epsilon_\perp^{\nu]\rho}S_{\perp\rho} \\
    & \times \sum_q e_q^2
    \bigg[ 2 \xbj g_T^q(\xbj) \sum_h \int d^2p_{hT} dz \, z\, D_1^{q,h}(z,p_{hT}) 
  + \frac{2\Lambda}{M} h_1^q(\xbj) \sum_h \int d^2p_{hT}  dz \, \tilde E^{q,h}(z,p_{hT}) \bigg] + \ldots \ , \nonumber
\end{align}
where $g_T^q(z,p_{hT})$ and $\tilde E^q(z,p_{hT})$ are twist-3 TMDs originating, respectively, from the quark-quark and the quark-gluon-quark fragmentation correlators. Note that in Eq.~\eqref{eq:Wsidis_ini} $M$ is the proton's mass, and we identified the power counting scale $M_h$ of Ref.~\cite{Mulders:1995dh} with our $\Lambda$. For clarity, we also reintroduced the quark flavors $q$, $e_q$ being their respective electric charge. 
The first term can be easily integrated with the help of the sum rules
\eqref{eq:SIDIS_to_DIS_TMDlevel1} and \eqref{eq:SIDIS_to_DIS_TMDlevel2}. To integrate the second term, however, we first need make
use of the relation $\tilde E(z) = E(z) - (\mq/\Lambda) z D_1(z)$, which is a
consequence of the QCD equations of motion \cite{Bacchetta:2006tn}, then
make again use of the sum rules \eqref{eq:SIDIS_to_DIS_TMDlevel1}-\eqref{eq:SIDIS_to_DIS_TMDlevel2} to obtain
\begin{align}
  \sum_h \int dz d^2p_{hT} \tilde E^{q,h}(z,p_{hT}) 
    = \xi_1 - \frac{\mq}{\Lambda} \xi_2 = \frac{\mj - \mq}{\Lambda }\ .
\end{align}
This formula provides us with a
nonperturbative generalization of the commonly used $\int\tilde E =0$ sum
rule introduced in \cite{Jaffe:1996zw}. Indeed, calculating the jet correlator 
on the perturbative vacuum one would obtain, as already discussed, $\mj=\mq$
and the integral would vanish.

Finally, the contraction of the hadronic tensor with the leptonic tensor leads
to the following result for the inclusive DIS cross section up to order $M/Q$~\cite{Bacchetta:2006tn}:
\begin{align}
\frac{d\sigma}{d\xbj \, dy\, d\phi_S}
=
\frac{2 \alpha^2}{\xbj y Q^2}\,
\frac{y^2}{2\,(1-\varepsilon)}\, 
\biggl\{
&F_{T} + \varepsilon F_{L}
+ S_\parallel \lambda_e\,
  \sqrt{1-\varepsilon^2}\; 
F_{LL}
+ |\bm{S}_\perp| \lambda_e\, \sqrt{2\,\varepsilon (1-\varepsilon)}\, 
  \cos\phi_S\, 
F_{LT}^{\cos \phi_S}
 \biggr\} \ ,
\label{e:crossdis}
\end{align}
where $\phi_S$ is the angle between the transverse component of the proton spin vector and the lepton plane, $\epsilon$ is the ratio of the longitudinal and transverse photon fluxes, and $\lambda_e$ is the electron's helicity. 
The structure functions on the right hand side read
\begin{align}
F_{T} &= \xbj\,\sum_q e_q^2\,f_1^q(\xbj),
\\
F_{L} &= 0,
\\
F_{LL} &=\xbj\,\sum_q e_q^2\,g_1^q(\xbj),
\label{e:FLLint}
\\
F_{UT}^{\sin \phi_S}&=0,
\label{e:FUTint}
\\
F_{LT}^{\cos \phi_S}&=-\xbj\,\sum_q e_q^2\, \frac{2M}{Q}\,
\biggl(\xbj  g_T^q(\xbj)
   + \frac{\mj -\mq}{M} \, h_{1}^q(\xbj) \biggr) \ ,
\label{e:FLTint}
\end{align}
where $f_1^q$, $g_1^q$ and $h_1^q$ are the unpolarized, polarized, and transversity PDFs,respectively.
The second term in the last structure function is a new result from our
analysis, it is proportional to the jet mass, and it is not suppressed as an
inverse power of $Q$ compared to the 
standard $g_T$ term. Perturbatively, $M_q^\text{pert}=m_q$ and the new term vanishes. However, on the nonperturbative vacuum the jet mass $M_q$ is much larger than the quark's current mass $m_q$, originating a nonnegligible term to the twist-3 part of the target's $g_2$ structure function, as we will discuss in the next section.

\section{The $g_2$ structure function}
\label{sec:g2}

The structure functions in Eqs.~\eqref{e:FLLint} -- \eqref{e:FLTint} can be
related to the usual structure functions $g_1$ and $g_2$ defined from the following Lorentz decomposition of the antysymmetric part of the inclusive hadronic tensor
\begin{align}
\begin{split}
W_A^{\mu\nu}(P,q) 
  & = \frac{1}{P\cdot q} \varepsilon^{\mu\nu\rho\sigma} q_\rho 
  \Big[ S_\sigma g_1(x_B,Q^2) 
    + \Big( S_\sigma - \frac{S\cdot q}{P\cdot q}\, p_\sigma
	\Big) g_2(x_B,Q^2)
  \Big] \ .
\label{eq:Wmunu}
\end{split}
\end{align}
Then, neglecting contrinutions of order $1/Q^2$, one obtains \cite{Bacchetta:2006tn},
\begin{align}
  g_1 &= \frac{1}{2x_B} F_{LL} \\
  g_2+g_1 &= - \frac{Q}{4 x_B^2 M} F_{LT}^{\cos \phi_S} \ .
\end{align}
Utilizing equations of motion and Lorentz invariance relations as discussed in Ref.~\cite{Accardi:2009au} to decompose $g_T$ into ``pure twist-3'' and twist-2 pieces, we arrive at
\begin{align}
\label{e:g2}
  g_2(\xbj) = g_2^{WW}(\xbj) + \frac{1}{2}\,\sum_a e_a^2
\biggl(
    \widetilde g_T^{a \star}(\xbj) 
    + \int_\xbj^1\frac{dy}{y} \widehat{g}_T^q(y) 
    + \frac{\mq}{M} \left(\frac{h_1^q}{x}\right)^\star(\xbj) 
    + \frac{\mj-\mq}{M} \frac{h_1^q(\xbj)}{\xbj} 
\Biggr) \ ,
\end{align}
where we used $f^*(x) = f(x) - \int_x^1\frac{dy}{y} f(y)$, and $\tilde g_T$ and $\hat g_T$ are pure twist-3 functions that only depend on projections of quark-gluon-quark correlator, and are explicitly defined in that reference. The first four terms coincide with the result obtained in the conventional handbag
approximation \cite{Accardi:2009au}, while the last is new. Note that even if
the relation is written for the sum over quark flavors weighted by their charge
squared, it is also valid flavor by flavor; in fact, the steps
leading to such a decomposition are formulated at the quark correlator level.

The first term is also known as the Wandzura-Wilczek function, $g_2^{WW} =
-g_1^*$, with $g_1=\frac12 \sum_q e_q^2 g_1^q$, and contains all the twist-2 chiral-even contributions to the $g_2$ structure coming from quark-quark correlators. The second and third
terms contain all ``pure twist-3'' contributions, i.e., those coming from
quark-gluon-quark correlators. The fourth and fifth terms contain chiral-odd twist-2 contributions and depend on the
transversity distribution function, $h_1$. 
The fourth term is usually neglected for
light quarks since it is proportional to $\mq={\cal O}$(1 MeV). The last term,
new in our analysis, is again proportional to the transversity distribution
but multiplied by the jet mass $\mj={\cal O}$(100 MeV), so that it cannot be a priori neglected.

\begin{figure}[tb]
\begin{center}
\includegraphics[width=8cm]{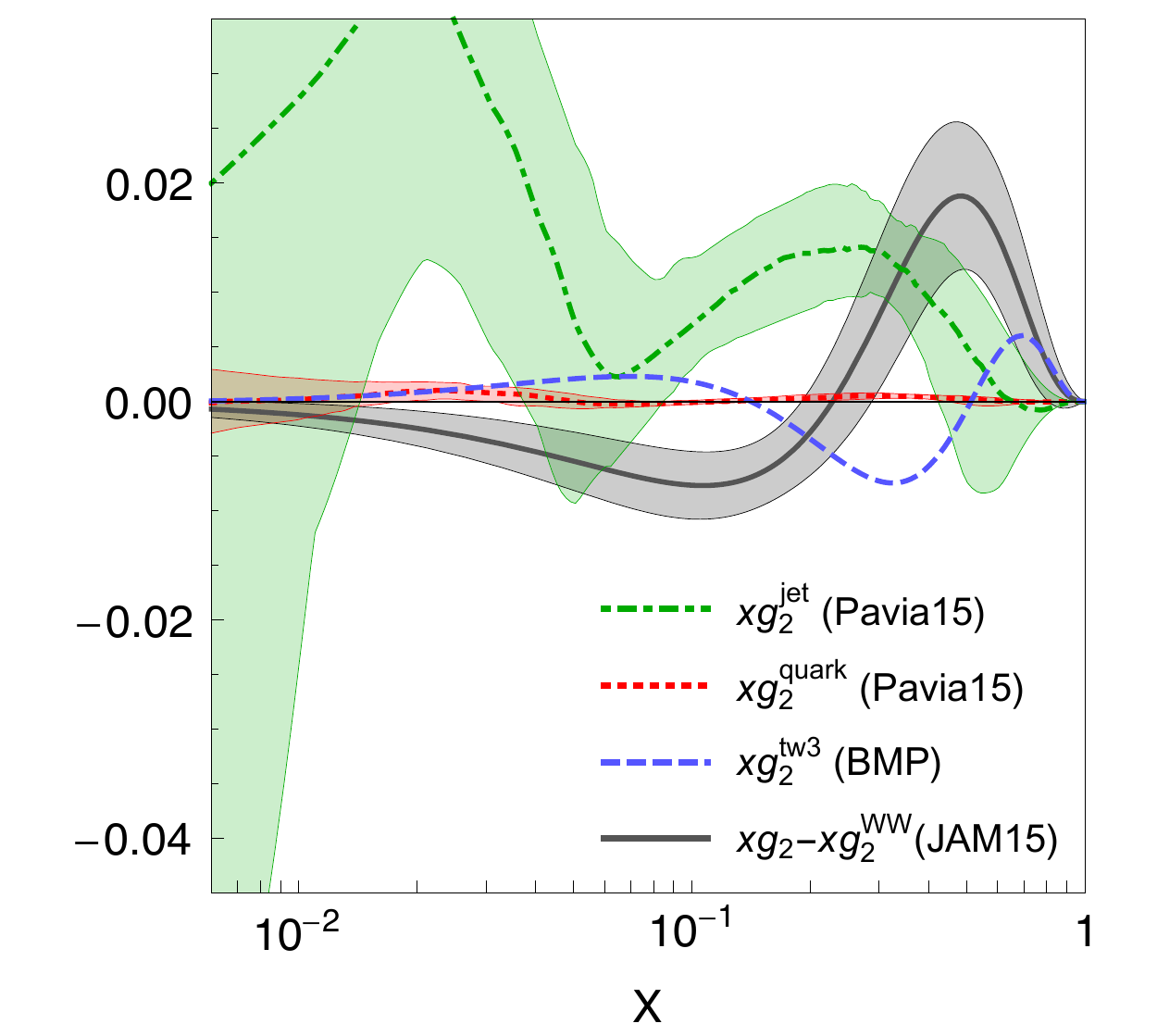}
\includegraphics[width=8cm]{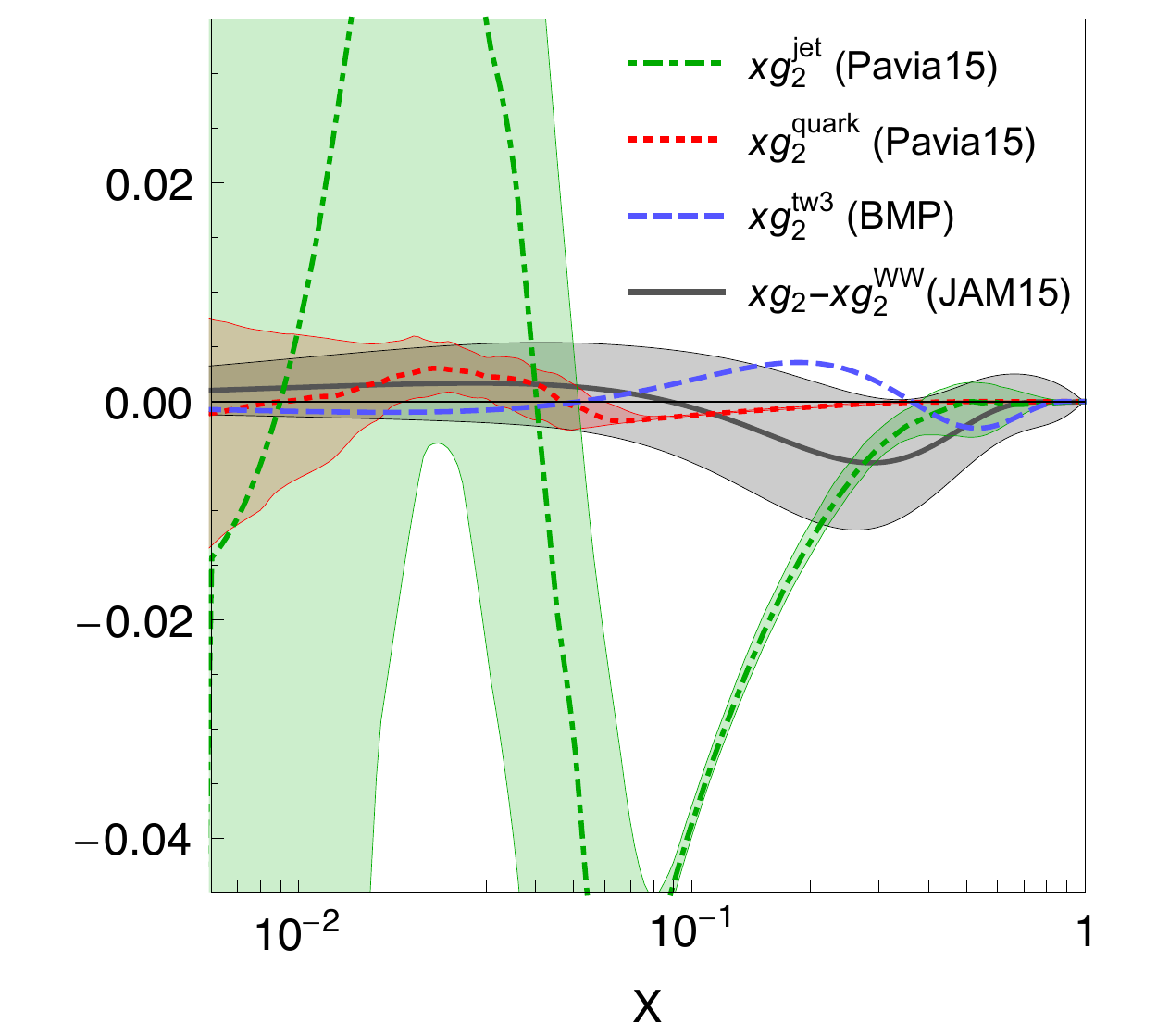}
\vskip-0.2cm
\caption{\label{f:g2contrib} 
Different contributions to the non Wandzura-Wilczek part of the proton (left)
and neutron (right) $g_2$ structure functions compared to the JAM15 fit of the
$g_2-g_2^{\text{WW}}$ function (solid black) \cite{Sato:2016tuz}. The quark and jet
contributions are shown with a dotted red and a dot-dashed green line
respectively, with uncertainty bands coming form the Pavia15 fit of the
transversity function \cite{Radici:2015mwa}. The uncertainty in the choice
$m_q=5$ MeV and $\mj=100$ MeV is not shown. The pure twist-3 contribution
calculated by Braun et al. \cite{Braun:2011aw} is shown as a dashed blue line
(no uncertainty estimate was provided in the original reference). 
}
\end{center}
\end{figure}

It is important to estimate the size of the various contributions to the non Wandzura-Wilczek part of $g_2$. We define the shorthand notation
\begin{align}
g_2^{\rm tw3} & = \frac{1}{2}\,\sum_q e_q^2
    \biggl(
    \widetilde g_T^{q \star}(\xbj) 
    + \int_\xbj^1\frac{dy}{y} \widehat{g}_T^q(y) 
    \biggr) 
\nonumber \\ 
g_2^{\text{quark}} &= \frac{1}{2}\,\sum_q e_q^2 
 \frac{\mq}{M} \bigg( \frac{h_1^q}{x} \bigg)^\star\!\!(\xbj) \ ,
\\
g_2^{\text{jet}} &= \frac{1}{2}\,\sum_q e_q^2 
\frac{\mj-\mq}{M} \frac{h_1^q(\xbj)}{\xbj} \ .
\nonumber 
\end{align} 
These terms are compared in Figure~\ref{f:g2contrib} to the $g_2-g_2^{WW}$ function obtained in the very recent JAM15 fit of polarized DIS asymmetries
\cite{Sato:2016tuz}, that includes a large amount of precise data at large $\xbj$ and small $Q^2$ from Jefferson Lab, and simultaneously fits the higher-twist components of
$g_1$ and $g_2$ to the data.\footnote{Notice, however, that the JAM15 fit imposes the $\int dx g_2(x) =0$ Burkhardt--Cottingham sum rule, which, however, is broken by inclusion of jet correlators, as discussed in Section~\ref{sec:moments}.} For the ``pure twist-3'' contribution,
$g_2^{\rm tw3}$, {\it i.e.}, the contribution from quark-gluon-quark matrix
elements, we show a recent light-front model calculation by Braun et al.~\cite{Braun:2011aw} (for bag model calculations, see \cite{Jaffe:1990qh,Stratmann:1993aw}). To estimate the contributions
from quark ($g_2^{\text{quark}}$) and jet mass ($g_2^{\text{jet}}$) effects, that depend on chiral-odd quark-quark matrix elements, we use the recent Pavia15 fit of the
transversity distribution from Ref.~\cite{Radici:2015mwa}, which is comparable
also to other 
extractions~\cite{Anselmino:2013vqa,Kang:2015msa}. Furthermore, we choose the
values of the mass parameters to be $\mq=5$ MeV and $\mj = 100$ MeV. 

As one can see, in the proton case the pure twist-3 contribution is quite
smaller in magnitude than, and nearly opposite in sign compared to, the twist-3 term extracted in the JAM15 fit. The quark-mass contribution, as expected, is essentially negligible. 
For what concerns the jet-mass contribution, the uncertainties due to the $h_1$
extraction are very large, especially at low $\xbj$. In addition, there is an
overall normalization uncertainty due to the choice of $\mj$, not shown in the
plot. In any case, the jet-mass contribution is strikingly large, and of the same order of magnitude as the chiral-even twist 3 term.

If we assume the latter to be of the order of the model calculation
by Braun et al., the breaking of the Wandzura-Wilczek relation can be used to
constrain the extractions of the transversity distribution. This is in particular true at low $\xbj$, where the pure twist-3 term is expected to vanish. 
Moreover, it is quite clear that the gap
between the pure twist-3 $g_2^{\rm tw3}$ function and the JAM15 fit can be
explained by the new jet-mass contribution we discuss in this paper.  

In the neutron case, the jet contribution is very negative at intermediate to
large values of $\xbj$. 
If one trusts the order of magnitude of the $g_2^{\rm tw3}$
calculation by Braun et al., one would conclude that the jet contribution
should not be that large. However, for a neutron target, $g_2^{jet}$ depends strongly on the $d$ quark's transversity, whose fit suffers from large systematic uncertainties and saturates the negative Soffer bound. Recent data in $p+p$ collisions indicate, in fact, that $h_1^{q=d}$ might be less negative than in the Pavia15 fits~\cite{Radici:2016lam}.
Correspondingly the jet contribution to the proton at $\xbj \approx 0.1$
would become less positive, improving as well the agreement with the JAM15
fit. 

\section{Moments of the $g_2$ structure function}
\label{sec:moments}

It is interesting to consider the moments of the non Wandzura-Wilczek contribution to $g_2$,
\begin{align}
  d_N \equiv (N+1) \int_0^1 dx\,x^N \bigg( g_2(x) - g_2^{WW}(x) \bigg) \ .
\end{align}
For a generic function $f$, let us define it's $N$-th moment as $f[N]=\int_0^1 dx\, x^{N} f(x)$. It is then straightforward to verify that $f^*[N] = f[N]\,N/(N+1)$ and  
\begin{align}
  d_N & = (N+1) g_2[N] + N g_1[N] \\
  & = \frac12 \sum_q e_q^2 \bigg[ N \tilde g_T^q[N] + \hat g_T^q[N]  
    + \frac{(N+1) \mj- \mq}{M} h_1^q[N-1] \bigg] \ .
\end{align}

The zero-th moment, $d_0=\int g_2$, provides an interesting relationship between transversity and the inclusive structure function $g_2$:
\begin{align}
  \label{eq:BC}
  \int dx\, g_2(x) = \sum_q e_q^2 \frac{\mj-\mq}{M} \int dx\, \frac{1}{x} h_1^q(x) \ .
\end{align}
Assuming $M_\text{light}\equiv M_u -m_u \approx M_d -m_d$ and dominance of light quarks in the sum over flavors, we can also write
\begin{align}
  \label{eq:BC-sfn}
  \int dx\, g_2(x) = \frac{M_\text{light}}{M} \int dx\, \frac{h_1(x)}{x} \ ,
\end{align}
with $h_1$ the transversity structure function.

In Eq.~\eqref{eq:BC}, we used the fact that $\widehat g_T^q[0]$ vanishes identically due to the symmetry properties of the quark-gluon-quark correlators \cite{Accardi:2009au}.
Therefore all pure twist-3 terms have explicitly disappeared, and the only surviving term on the right-hand side is the new jet contribution.
Thus, our new sum rule \eqref{eq:BC}
generalizes the Burkhardt--Cottingham (BC) sum rule~\cite{Burkhardt:1970ti},
which states that  $\int_0^1 dx\, g_2(x) =0$, while we have shown that jet-mass corrections, and in particular from invariant mass generation in spin-flip processes, can directly violate this.
In fact, the possibility of a violation of the BC sum rule due to 
contributions from spin-flip processes was already mentioned in the original
derivation, but these do not show up in treatments that only
consider free-field quark propagators for the struck quark
\cite{Jaffe:1996zw}.  Although we formulated \eqref{eq:BC} in terms of a sum over quark flavors in order to display a clear
connection to the structure function $g_2$, we stress that this is valid also
flavor by flavor, i.e., for each single flavor the only measurable nonzero contribution to the zeroth moment of the
structure function $g_2^q$ comes from the coupling between its jet mass and transversity function\footnote{This conclusion is true even if the BC sum rule is broken by a $J = 0$ fixed pole with
non-polynomial residue \cite{Jaffe:1996zw}, since this would appear as a
$\delta(\xbj)$ contribution and would not be measurable.}.

One should notice that since $h_1$ is slowly driven to 0 by QCD evolution as $Q^2 \to \infty$, the BC sum rule may still be satisfied at least asymptotically. At finite scales, however, the only way to preserve the validity of the BC sum rule is if
$
%\begin{align}
   \int dx\, \frac{1}{x} h_1^q(x) = 0 \ .
\label{eq:ABsumrule}
%\end{align}
$
Interestingly, one can show that this constraint, if valid at any given scale
$Q_0$, is conserved through QCD evolution. However, we think that it is unlikely to be satisfied in general, since the right hand side is different from zero
in perturbative QCD~\cite{Kundu:2001pk}, as well as in model calculations \cite{Schweitzer:2001sr,Wakamatsu:2007nc,Pasquini:2005dk,Cloet:2007em,Bacchetta:2008af,Bourrely:2010ng}.
A finite breaking of the BC sum rule would imply that $h_1(x)/x$ must be integrable, which is possible only if, at small $x$, the transversity goes as $h_1^q(x) \propto x^\epsilon$ with $\epsilon>0$. While $\epsilon=1$ in perturbative QCD \cite{Hayashigaki:1997dn,Vogelsang:1997ak}, the leading Regge contribution at small $x$ indicates that $\epsilon = 0$  \cite{Kirschner:1996jj}, and opens the door to a more drastic breaking of the sum rule.
Finally, we note that the small-$x$ behavior of the longitudinal spin structure function $g_1$ has been recently studied in Ref.~\cite{Kovchegov:2016zex}; however, since the small $x$ structure of the $h_1$ function may be quite different from that of $g_1$ \cite{Kirschner:1996jj},
it would be interesting to extend those techniques to the transverse spin structure functions $g_T=g_1+g_2$ and $h_1$, and investigate their role in the breaking of the BC sum rule.

The first moment, $d_1$, is the first one to display a contribution from the pure twist-3 part of $g_2$:
\begin{align}
  d_1 & = \frac12 \sum_q e_q^2 \bigg( 2 \tilde g_T^q[1] + \hat g_T^q[1] 
    + \frac{2\mj-\mq}{M} h_1^q[0] \bigg)
\label{e:d1}
\end{align}
where $h_1^q[0] = \int_0^1 dx h_1^q(x)$ is the contribution of a quark $q$ to the target's tensor charge. The second moment,
\begin{align}
  d_2 & = \frac12 \sum_q e_q^2 \bigg( 3 \tilde g_T^q[2] + \hat g_T^q[2] 
    + \frac{3\mj-\mq}{M} h_1^q[1] \bigg) \ ,
\end{align}
is also interesting because the pure twist-3 part can be related to quark-gluon-quark local matrix elements, see \cite{Jaffe:1996zw}, and
interpreted as the average color force experienced by the struck quark as
it exits the nucleon \cite{Burkardt:2012sd}; for experimental
measurements of $d_2$, see, e.g., Refs.~\cite{Anthony:2002hy,Slifer:2008xu,Solvignon:2013yun,Posik:2014usi,Flay:2016wie}. 

For both the $d_1$ and $d_2$ moments, the transversity contribution is a background to the extraction of the pure twist-3 piece. Fortunately, it is a quantity that can be extracted from the lattice
\cite{Green:2012ej,Bali:2014nma,Bhattacharya:2015wna,Abdel-Rehim:2015owa,Bhattacharya:2016zcn} 
or extracted form experimental data \cite{Radici:2015mwa,Anselmino:2015sxa,Kang:2015msa},
and information from the extended BC sum rule \eqref{eq:BC-sfn} promises to improve future transversity fits. Furthermore, as combined QCD fits of different distribution functions have now become possible \cite{Ethier:2017zbq}, the jet mass $M_q$ could also be considered as a free parameter in a combined helicity and transversity PDF fit. Therefore the pure twist-3 part can, in principle, be properly isolated and measured.

\begin{figure}[tbh]
  \centering
  \parbox[c]{8cm}{\vskip.1cm\centering
  \includegraphics
      [width=7cm,clip=true]
      {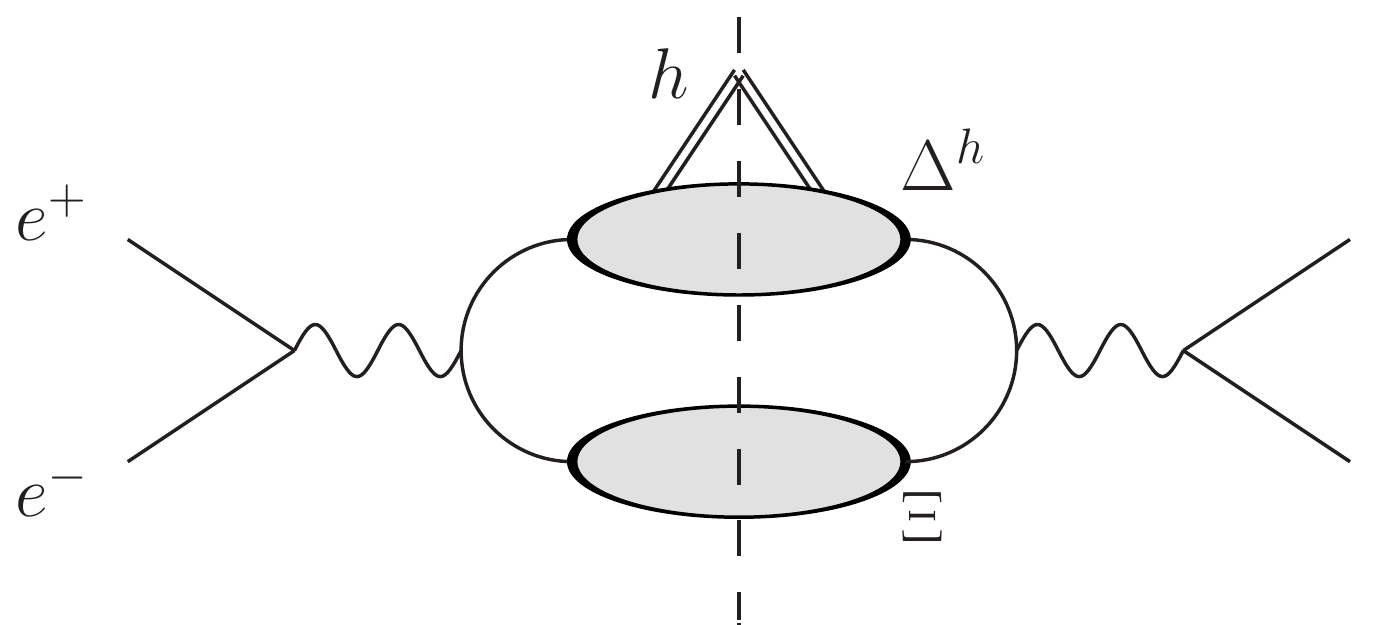}
  }
  \parbox[c]{8cm}{\vskip-0.2cm\centering
  \includegraphics
      [width=7cm,clip=true]
      {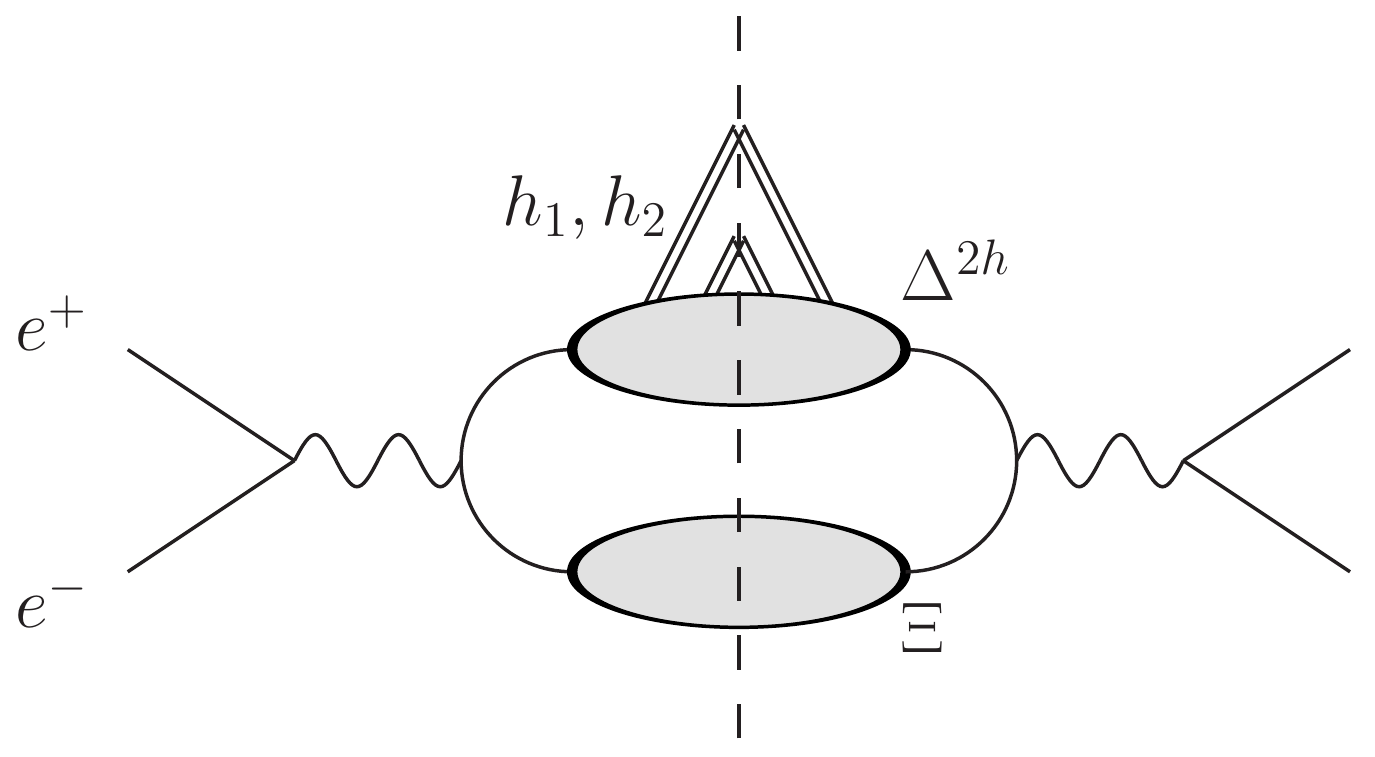}
  }
  \caption{\small
	Single hadron {\it (left)} and double hadron {\it (right)}
	production in $e^+ e^-$ collisions at LO with jet and
	fragmentation correlators.
  }
  \label{fig:epl_emn_h}
\end{figure}

We should also note that the $\mj$ jet mass parameter can
be experimentally measured, \eg, in electron-positron collisions.
A promising avenue is through inclusive single hadron production, $e^+ e^- \to h X$, and inclusive dihadron 
production from the same hemisphere, $e^+ e^- \to h h X$, see~Fig.~\ref{fig:epl_emn_h}.
In single-hadron production, the fragmentation functions 
play the role of PDFs in DIS and couple to the jet functions
in an analogous way. To access the spin-flip $J_1$ function one needs to detect a polarized hadron, such as a $\Delta$ baryon. 
In double hadron production, the enlarged number of Dirac structures
of the dihadron fragmentation correlators related to the relative momentum of the two hadrons \cite{Bacchetta:2002ux,Bacchetta:2003vn} allows one to access
the jet function in novel ways, and in particular to isolate the
contribution from the helicity-flip $J_1$ term in combination with the
chiral-odd fragmentation function $H_1^{\sphericalangle}$. 

To conclude this section, we note that the jet contribution also leads to an explicit breaking of the Efremov--Teryaev--Leader (ETL) sum rule \cite{Efremov:1996hd}, in which the pure twist-3 contribution to the first moment of $g_2-g_2^{WW}$ also disappears. To see this, let's define the valence contribution to a given structure function as $f^V=\frac12\sum_q e_q^2 (f^q-f^{\bar q})$. Then, as shown in \cite{Efremov:1996hd}, $\widehat 2\tilde g_T^V[1] + g_T^V[1]  = 0$, and from Eq.~\eqref{e:d1} we obtain
\begin{align}
  d_1^V = \frac12 \sum_q e_q^2 \frac{2\mj- \mq}{M} 
    \big(h_1^{q}[0]-h_1^{\bar q}[0]\big) \ .
\end{align}
Assuming again dominance of light flavors, we can also see that 
\begin{align}
  d_1^V = \frac{M_{\text{light}}}{M} \delta_T(p) \ ,   
\end{align}
This gives an alternative way to access the proton tensor charge, $\delta_T(p) = \sum_q e_q^2 \big(h_1^{q}[0]-h_1^{\bar q}[0]\big)$, by measuring or fitting moments of the flavor separated $g_2$ structure function.

%%%%%%%%%%%%%%%%%%%%%%%%%%%%%%%%%%%%%%%%%%%%%%%%%%%%%%%%%%%%%%%%%%%%%%%%%
\section{Conclusions}

In this paper, we revisited the inclusive DIS analysis, including the
effects due to the production of a system of final state hadrons in the current direction, which we
conveniently referred to as a ``jet.'' We described this in terms of a jet
correlator that corresponds, up to twist-4 contributions, to the nonperturbative quark cut propagator, or, equivalently, to the quark's spectral function, and of a quark-gluon-quark jet correlator needed to insure gauge invariance of the calculation.
We then carried out the analysis of
the DIS cross section up to contribution of order $1/Q$. 
The introduction of the jet correlators 
leads to a difference in the expression of the structure function $g_2$ in
inclusive DIS with respect to the standard analysis: a new term appears, proportional to a jet mass parameter $\mj={\cal O}$(10-100 MeV)
and to the transversity distribution function. This new term
contributes to the violation of the Wandzura-Wilczek relation, in addition to
the standard pure twist-3 terms and quark mass corrections. Contrary to these
standard terms, however, the new jet mass correction does not necessarily integrate to zero and so violates also the
Burkhardt--Cottingham and Efremov--Teryaev--Leader sum rules. This is yet another example of how surprising
and rich the phenomenology of polarized inclusive DIS can be,
and offers a new direction for theoretical studies and experimental investigations of spin physics over a wide range in $x$, from the valence and sea regions at Jefferson Lab \cite{Dudek:2012vr} to the small-$x$ region at the future Electron-Ion Collider \cite{Accardi:2012qut}.

Detailed measurements of the $g_2$ structure
function can be used to constrain the jet mass parameter $M_q$, the transversity
distribution function and the nucleon tensor charge, helping their extraction from other observables, e.g., in electron-positron annihilation and semi-inclusive DIS. Knowledge of the jet mass parameter and of the transversity distribution will eventually be needed for a precise extraction of pure twist-3 terms from the $g_2$ structure function, or from transverse target single spin asymmetries \cite{Schlegel:2012ve}.

Finally, studying and classifying all the contributions of jet correlators to single and double hadron production in electron-positron annihilation events will open up a rich phenomenology. Measurements in the asymptotically large $Q^2$ regime will provide access to the integral of the $J_1$ jet function, i.e., to the jet-mass parameter $\mj$, and therefore (in conjunction with precise measurements or lattice QCD calculations of the first $h_1$ moment) also of the target's tensor charge through the modified ETL sum rule. 
Equally interesting is the possibility to experimentally measure, at finite values of $Q^2$, the momentum dependence of the jet functions $J_1$ and $J_2$, that enter structure functions integrated only up to $\sigma^2=Q^2(1/x_B-1)$~\cite{Accardi:2008ne}. In other words, it may become possible to experimentally access also the quark's spectral function itself.

%%%%%%%%%%%%%%%%%%%%%%%%%%%%%%%%%%%%%%%%
%%%%%%%%% ACKNOWLEDGMENTS %%%%%%%%%%%%%%
%%%%%%%%%%%%%%%%%%%%%%%%%%%%%%%%%%%%%%%%

\begin{acknowledgments}
We are grateful to A.~Signori for a careful reading of a draft of this article, and to A.~Mukherjee, M.~Stratmann, and W.~Vogelsang for interesting and informative discussions. This work has been supported by the United States Department of Energy (DOE) contract No. DE-AC05-06OR23177,
under which Jefferson Science Associates, LLC operates Jefferson Lab, by the DOE contract DE-SC008791, and 
by the European Research Council (ERC) under the European Union's 
Horizon 2020 research and innovation programme (grant agreement No. 647981,
3DSPIN)
\end{acknowledgments}

\bibliographystyle{myrevtex}
\bibliography{jetbiblio}

\end{document}